\documentclass[letterpaper]{article}

\usepackage[round]{natbib}

\usepackage{fullpage}  
\usepackage{natbib} 
\usepackage{authblk} 

\usepackage{graphicx}
\usepackage{amsthm}
\usepackage{booktabs} 
\usepackage[ruled]{algorithm2e} 

\SetAlFnt{\small}
\SetAlCapFnt{\small}
\SetAlCapNameFnt{\small}
\SetAlCapHSkip{0pt}
\IncMargin{-\parindent}

\usepackage{amsmath}
\usepackage{booktabs}
\usepackage{float}

\setcitestyle{authoryear}

\newcommand{\ds}{\displaystyle}
\newcommand{\reals}{\mathbf{R}}

\newcommand{\Nj}{N^{-j}}

\newcommand{\vp}{\mathbf{p}}
\newcommand{\vk}{\mathbf{k}}
\newcommand{\vK}{\mathbf{K}}

\newcommand{\Primal}{\mbox{P}}
\newcommand{\Dual}{\mbox{D}}
\newcommand{\UCEPrimal}{\tilde{\Primal}}
\newcommand{\UCEDual}{\tilde{\Dual}}

\newcommand{\mK}{\mathcal{K}}
\newcommand{\vcg}{g}
\newcommand{\mindemand}{\underline{\kappa}}
\newcommand{\maxdemand}{\overline{\kappa}}
\newcommand{\capacity}{\gamma}
\newcommand{\bindingecon}{E_i(\vk)}
\newcommand{\prices}{\boldsymbol{\rho}}
\DeclareMathOperator*{\argmax}{arg\,max}
\DeclareMathOperator*{\argmin}{arg\,min}

\theoremstyle{definition}
\newtheorem*{definition}{Definition}
\newtheorem{theorem}{Theorem}

\newtheorem{lemma}[theorem]{Lemma}

\title{Iterative Vickrey Auctions via Linear Programming}

\author[1]{S\'{e}bastien Lahaie}
\author[2]{Benjamin Lubin}
\affil[1]{Google Research}
\affil[2]{Boston University}
\date{July 14, 2026}

\begin{document}

\maketitle

\begin{abstract}
Building on the linear programming approach to competitive equilibrium pricing, we develop a general method for constructing iterative auctions that achieve Vickrey-Clarke-Groves (VCG) outcomes. We show how to transform a linear program characterizing competitive equilibrium prices into one that characterizes universal competitive equilibrium (UCE) prices, which elicit precisely the information needed to compute VCG payments. By applying a primal-dual algorithm to these transformed programs, we derive iterative Vickrey auctions that maintain a single price path, eliminating the overhead and incentive problems associated with multiple price paths used solely for payment calculations. We demonstrate the versatility of our method by developing a novel iterative Vickrey auction for the multi-unit setting and an iterative variant of the Product-Mix auction. The resulting auctions combine the transparency of iterative price discovery with the efficiency and incentive properties of the VCG mechanism.
\end{abstract}


\section{Introduction}
\label{sec:introduction}

The Vickrey-Clarke-Groves (VCG) mechanism plays a central role in auction theory due to its efficiency and incentive properties. In its traditional formulation, the mechanism is a sealed-bid auction: agents report their valuations to a center who implements an efficient allocation and charges payments that incentivize truthful bidding. In practice, however, iterative auction formats offer several advantages over their sealed-bid counterparts: they reveal less private information, decrease communication costs, and often have more transparent and intuitive rules~\citep{rothkopf1990vickrey,ausubel2004efficient,ausubel2006lovely}. Based on the well-known equivalence between the Vickrey and English auctions for a single item, there has therefore been substantial research into iterative versions of the VCG auction in more general settings with multiple homogeneous or heterogeneous items~\citep{de2007ascending,demange1986multi,mishra2009multi,gul2000english}.

Implementing VCG payments requires solving $n+1$ separate optimization problems for $n$ bidders---one for the main economy with all bidders, and one for each marginal economy where a single bidder is excluded. Mirroring this structure, \citet{ausubel2006efficient} develops an iterative auction for agents with substitutes valuations that maintains $n+1$ price paths (i.e, agents bid in response to different prices in parallel), and combines their information to calculate VCG payments. This idea is further generalized by~\citet{de2007ascending} to agents with general valuations. However, the use of multiple price paths requires each agent to bid on paths that are only used to calculate the payment of \emph{another} agent, with no other purpose. This is problematic in practice because agents have no reason to put any effort into bidding accurately on these price paths.
In this paper, we develop a general framework for constructing iterative, \emph{single price path} VCG auctions based on linear programming. Our approach builds on the insight that VCG payments can be derived from \emph{universal competitive equilibrium} (UCE) prices---prices that simultaneously clear both the main economy and all marginal economies~\citep{parkes2002price}. These prices can be used to drive forward the bidding process towards an efficient allocation, but also contain precisely the information needed to compute VCG payments~\citep{mishra2007ascending}. Our central contribution is to show how one can transform a linear program (LP) for competitive equilibrium prices into an LP that characterizes UCE prices. This transformed program can then be solved using a primal-dual algorithm, following a methodology pioneered by~\citet{bikhchandani2002linear}, leading to iterative auctions that converge to VCG outcomes using a single price path.
Importantly, our approach preserves key economic properties of the original LP and its dual such as integrality (to properly handle indivisible items), and the structures of the CE and UCE prices are closely related---simpler CE prices (e.g., linear and anonymous) lead to simpler UCE prices. Specifically, our UCE program computes \emph{lower envelope} prices: it takes the clearing prices from the main and marginal economies, adds suitable agent- and economy-specific offsets, and forms UCE prices as the minimum of these affine functions. If the original CE prices have dimension $d$, the UCE prices have dimension $n(n + d)$.
For homogeneous items and agents with decreasing marginal values, UCE prices have dimension $n(n + 1)$, but can be simplified to $nm$ where $m$ is the number of units. For heterogeneous items and substitutes valuations, our approach yields an auction with UCE prices of dimension $n(n + m)$, matching the communication requirements of~\citet{ausubel2006efficient}'s multi-path auction, but using a single price path.

Although we build on the linear programming framework for iterative VCG auctions of~\citet{bikhchandani2002linear} and~\citet{de2007ascending}, our UCE framework is arguably more straightforward. According to their recipe, the efficient allocation problem is first formulated as an LP, such that dual variables for this program characterize not just CE prices but also the agent's VCG payoffs. Our approach dispenses with the latter requirement. For heterogeneous items, the requirement can only be satisfied by an LP of polynomial size (in the number of items) if agents have unit-demand valuations~\citep{demange1986multi,leonard1983elicitation}. For agents with gross substitutes valuations and beyond, the LPs that have been proposed all result in personalized prices over all possible item bundles (the most expressive prices possible), but still rely on an agents-are-substitutes condition to characterize VCG payoffs~\citep{bikhchandani2002linear,de2007ascending,gul2000english,bikhchandani2006ascending}. A notable exception is the auction of~\citet{mishra2007ascending}, which also uses UCE prices, but in their work these are always represented as personalized bundle prices, with no simplifications offered for basic settings like homogeneous items.

To demonstrate how our approach leads to novel iterative Vickrey auctions in specific applications, we consider settings where the seller can allocate up to $K$ item units in aggregate, and where agents have constant or decreasing marginal values for additional units. 
With just a single item, our approach leads to an alternative to Ausubel's celebrated \emph{clinching auction}~\citep{ausubel2004efficient}. Although the auctions are equivalent in that they both implement the VCG outcome, the mechanics of our primal-dual auction are quite different. It quotes lower envelope prices, and bidders respond with the number of units they demand at these prices. The auction then finds an economy (main or marginal) that has not yet cleared based on these demands, and increments its price. For the other economies, the agent-specific additive offsets are incremented by the size of the smallest bundle in the agent's demand set.
When there are several types of items, and the cost differences to produce item types are constant on the part of the seller, our approach leads to the first (to our knowledge) iterative VCG auction for product-mix auctions that uses a single price path~\citep{klemperer2010product}.
We illustrate the mechanics of our auction on a worked example with multiple homogeneous units and agents with decreasing marginal valuations, and also revisit a stylized example of a spectrum auction from~\cite{ausubel2004efficient} to emphasize that our auction computes VCG payments without full information revelation.
We next evaluate our UCE auction on an instance of a product-mix auction for the 
allocation of loans, constructed based on summary statistics from Bank of England bidding data from~\cite{grace2024competing}. We compare against a benchmark that only computes CE prices (i.e, without enough information to compute VCG payments) and a benchmark that uses multiple price paths. We find that under ascending versions of the auctions, the UCE auction uses the same number of rounds as the CE price auction, and uses an order of magnitude fewer demand queries to bidders than the multiple price path auction. Under descending versions, the UCE auction typically uses 11\% more rounds to reach the VCG outcome beyond the linear price auction.

\textit{Incentives.} Our contribution is computational in nature. Iterative VCG auctions inherit the strong incentive properties of the sealed-bid VCG mechanism~\citep{de2007ascending}. By the Revelation Principle, if an iterative auction implements the VCG outcome, bidding according to one's true valuation dominates bidding according to a different valuation. However, without additional restrictions it is possible for an agent's bid responses to be inconsistent with \emph{any} valuation function. To rule out this kind of bidding behavior, the auction can apply activity rules, specifically revealed-preference activity rules~\citep{ausubel2020revealed,ausubel2014market,ausubel2017practical}. These rules only rely on the prices of demanded bundles at each round, not the actual price structure, and therefore can be seamlessly integrated into iterative UCE auctions.

\textit{Related Work.} Besides the literature on linear programming characterizations of VCG payments and primal-dual auctions already covered, our work relates to other strands of research. It connects to the literature on competitive equilibrium with discrete items (i.e., Walrasian equilibrium), including the fundamental work of~\citet{kelso1982job} on substitutes valuations, and earlier iterative auctions to allocate discrete items~\citep{demange1986multi,bikhchandani1997competitive,milgrom2000putting,gul1999walrasian,gul2000english}. Recent work on the design and analysis of the product mix auction also falls in this area, as the auction was designed with the substitutes property in mind~\citep{klemperer2010product,baldwin2019understanding,baldwin2024implementing,baldwin2024solving}.
Our work also connects with the broad literature on iterative combinatorial auctions~\citep{parkes1999bundle,parkes2006mit,ausubel2002ascending,parkes2000iterative}. Particularly relevant is research on the benefits of different price structures. \citet{kwasnica2005new} design a combinatorial auction with approximate linear clearing prices, finding informational benefits in lab experiments. In computational evaluations, \citet{bichler2009computational} and~\citet{schneider2007robustness,schneider2010robustness} find that simpler linear price structures (versus bundle prices) are more robust to deviations from sincere bidding, and have faster convergence in primal-dual auctions; these findings are also borne out in lab experiments~\citep{scheffel2011experimental}.

\textit{Outline.} Section~\ref{sec:model} specifies the model and formally defines the concept of UCE prices. Section~\ref{sec:lp-formulations} provides our transformation from an LP for CE prices to an LP for UCE prices. Section~\ref{sec:primal-dual-auction} applies the primal-dual algorithm to the UCE LP and presents the resulting iterative VCG auction for the multi-unit setting. 
Section~\ref{sec:applications} illustrates and evaluates our primal-dual UCE auction on instances of multi-unit, stylized spectrum and product-mix auctions.
Section~\ref{sec:conclusion} concludes.

\vspace{-5pt}
\section{The Model}
\label{sec:model}

We consider a general combinatorial auction model with a set of agents each interested in buying a set of items from a single seller. There are possibly multiple units of each item, and items are indivisible. Let $N$ be the set of agents, indexed by $i = 1, \dots, n$ where $n = |N|$. A \emph{bundle} $\vk$ represents a set of items, and an \emph{allocation} $\vK = (\vk_1, \dots, \vk_n)$ indicates the bundle that each agent receives. We let $\mK_i$ denote the \emph{consumption set} of agent $i$, i.e., the set of bundles it can acquire, and $v_i : \mK_i \rightarrow \reals_+$ its \emph{valuation} mapping bundles to values. We assume that each $\mK_i$ contains the zero bundle, denoted $\mathbf{0}$, and that valuations are normalized to $v_i(\mathbf{0}) = 0$. The seller's \emph{production set} consists of all feasible allocations it can form given the supply of items, denoted $\mK \subseteq \mK_1 \times \dots \times \mK_n$.

The \emph{prices} for the bundles are in general represented via mappings $\rho_i : \mK_i \rightarrow \reals_+$, expressing that the price for bundle $\vk$ to agent $i$ is $\rho_i(\vk)$. Note that prices can be personalized (i.e., agent-specific) in general.  In specific settings, prices typically have a particular structure (e.g., additive over items), and we will shortly refine our setup to accomodate price structure. Agents have quasi-linear utility: if agent $i$ acquires bundle $\vk$ at prices $\rho_i$, its utility is $v_i(\vk) - \rho_i(\vk)$. Agent $i$'s \emph{demand set} at prices $\rho_i$ is defined as $D_i(\rho_i) = \argmax_{\vk \in \mK_i} v_i(\vk) - \rho_i(\vk)$, namely the set of bundles that maximize its utility; we stress that this is a set-valued mapping, not just a selection of a single bundle.

We will refer to a subset of agents as an \emph{economy}. For an agent $i$, we write $N^{-i} = N \setminus \{i\}$. The economy $N^{-i}$ is the \emph{marginal economy} with agent $i$ removed, and $N$ is the \emph{main economy}. For notational convenience, we introduce a special index $0$ which is not associated to any agent, and we let $N^{-0} = N$. We let $N_0 = N \cup \{0\}$ and use $j = 0, 1, \dots, n$ to index the main and marginal economies.

Note that items are implicit in our model formulation; they are inherent in the structure of the production set $\mK$, which encodes constraints such as: no unit can be allocated to more than one agent (due to indivisibility). Because of this underlying structure, we can assume that for a feasible allocation $(\vk_1, \dots, \vk_n) \in \mK$, replacing any single bundle with $\mathbf{0}$ results in a feasible allocation; in particular, a feasible allocation in the corresponding marginal economy. The seller's \emph{supply set} at prices $\prices = (\rho_1, \dots, \rho_n)$ in economy $j \in N_0$ is defined as the set of feasible allocations that maximize the seller's revenue:
\begin{equation}
\label{eq:rev-max-obj}
S^{-j}(\prices) = \argmax_{\vK \in \mK} \left\{ \sum_{i \in N^{-j}} \rho_i(\vk_i) : \mbox{$\vk_j = \mathbf{0}$ $(j \neq 0)$} \right\}.
\end{equation}
In an iterative auction, prices are updated in order to balance supply and demand. We next recall the classic VCG mechanism, which is a single-shot auction, and then explain how an iterative pricing mechanism can implement its outcome.

\subsection{VCG Mechanism}

The VCG mechanism in our setting is a sealed bid auction where the agents first communicate their valuations $v_i$ to the seller. The seller selects an \emph{efficient} allocation, which maximizes
\begin{equation}
\label{eq:efficient-obj}
V(N) = \max_{\vK \in \mK} \: \sum_{i \in N} v_i(\vk_i).
\end{equation}
Next, the auction computes and charges a \emph{VCG payment} to each agent $i \in N$, defined as:
\begin{equation}
\nonumber
    \vcg_i = v_i(\vk_i) - [V(N) - V(N^{-i})].
\end{equation}
Note that VCG payments are not unique, because they depend on the choice of the efficient allocation (which itself may not be unique). On the other hand, an agent's VCG payoff $v_i(\vk_i) - \vcg_i = V(N) - V(N^{-i})$ is fixed and does not depend on the selected allocation. Agents are not required to communicate their actual true valuations to the seller. However, it is a classic result that truthful reporting of a valuation is a dominant strategy equilibrium in the VCG mechanism~\citep{krishna2009auction}, so we have described it assuming truthful communication of each $v_i$.
Note that VCG payments are not prices: they simply define a charge for each agent based on the selected efficient allocation, whereas prices are defined over all possible bundles an agent might acquire. Comparing~\eqref{eq:rev-max-obj} to~\eqref{eq:efficient-obj}, we note an analogy between the general personalized bundle prices we have considered so far and agent valuations. However, in specific applications like multi-unit auctions and product-mix auctions, the final price structure will typically differ from structure found in agent valuations.

\subsection{UCE Prices}

A crucial link between VCG payments and prices was discovered by~\citet{parkes2002price}, who defined the concept of a universal competitive equilibrium. The following definitions are central to this paper.
\begin{definition}
For an economy $j \in N_0$, prices $\prices$ are \emph{competitive equilibrium} (CE) prices if there is a feasible allocation $(\vk_1, \dots, \vk_n) \in S^{-j}(\prices)$ such that $\vk_i \in D_i(\rho_i)$ for $i \in N^{-j}$. We say that the prices \emph{support} the allocation if these conditions hold. Prices $\prices$ are \emph{universal competitive equilibrium} (UCE) prices if they are simultaneously CE prices for all economies $j \in N_0$.
\end{definition}
By the first and second fundamental theorems of welfare, if prices $\prices$ support an allocation, then the allocation is efficient, and furthermore the prices support every possible efficient allocation~\citep{varian1992microeconomic}.
To elaborate on the definition, if $\prices$ are UCE prices, then they support any choice of efficient allocation $(\vk_1, \dots, \vk_n)$ in the main economy. Furthermore, taking any efficient allocation $(\vk^{-j}_1, \dots, \vk^{-j}_n)$ in marginal economy $j \in N$, the prices $\prices$ support this allocation as well. The important aspect here is that the efficient allocations might (necessarily) differ in the various economies, but the prices support all of them without the need to change the prices in any way. The following result was proved in~\cite{parkes2002price, mishra2007ascending}. We provide a proof in Appendix~\ref{app:omitted-proofs} for completeness.%
\begin{theorem}{\citep{parkes2002price}}
\label{thm:uce-vcg-construction}
Assume $(\vk_1, \dots, \vk_n)$ is an allocation in the main economy and $\prices$ are supporting UCE prices. Let $(\vk^{-j}_1, \dots, \vk^{-j}_n) \in S^{-j}(\prices)$ for $j \in N$ be revenue-maximizing allocations in the marginal economies at these prices. Then the VCG payment for agent $i$ with respect to $(\vk_1, \dots, \vk_n)$ is:
\begin{equation}
\label{eq:uce-vcg-formula}
\vcg_i = \sum_{\ell \in N^{-i}} \rho_{\ell}(\vk^{-i}_{\ell}) - \sum_{\ell \in N^{-i}} \rho_{\ell}(\vk_{\ell}).
\end{equation}
\end{theorem}
Therefore, for an iterative auction to implement VCG payments, it suffices to converge to UCE prices and an efficient allocation supported by these prices.
The revenue-maximizing allocations $(\vk^{-j}_1, \dots, \vk^{-j}_n) \in S^{-j}(\prices)$ in the marginal economies only serve to evaluate the optimal revenue in these economies, which appears as the first term in~\eqref{eq:uce-vcg-formula}.

Although UCE prices allow one to compute VCG payments, they do not contain enough information to recover the agent's full VCG payoffs. In this sense, an auction that converges to UCE prices elicits less information about valuations than the original single-shot VCG mechanism, which requires full revelation.

\subsection{Allocation and Price Structure}

To develop concrete iterative auctions for specific settings, we need to refine our general model to specify the structure of feasible allocations and prices. In their seminal paper, \citet{bikhchandani2002package} introduce several LP formulations for the allocation problem in combinatorial auctions, ranging from item pricing, to bundle pricing, to fully nonlinear and personalized bundle prices. Our framework for developing iterative Vickrey auctions applies to any of these. To cover all formulations at once, we introduce \emph{embeddings} that map bundles to a vector representation of the resources they contain. Prices are then linear functions over these embeddings.

Specifically, we let $\phi_i : \mK_i \rightarrow \reals^d$ be an embedding for agent $i$, mapping each bundle $\vk \in \mK_i$ to a $d$-dimensional vector $\phi_i(\vk)$. Prices are then represented via a price vector $p \in \reals^d$, such that the price for bundle $\vk$ to agent $i$ is given by the linear function $\rho_i(\vk) = p \cdot \phi_i(\vk)$. The demand set at prices $p$ is therefore $D_i(p) = \argmax_{\vk \in \mK_i} v_i(\vk) - p \cdot \phi_i(\vk)$.
Depending on the choice of embedding $\phi$, this framework can encode a variety of price structures:
\begin{itemize}
    \setlength{\itemsep}{0.5em}
    \item \textit{Linear.} If items are indexed by $m = 1, \dots, M$ and there are multiple units of each item, we can take $d = M$ and let $\phi_i(\vk)$ be the vector of item quantities in bundle $\vk$, so that $\phi_i(\vk) = (k_1, \dots, k_M)$ where $k_m$ is the number of units of item $m$ in bundle $\vk$. Prices are then linear over items, with $p = (p_{1}, \dots, p_{M})$ representing the per-unit prices for each item.
    \item \textit{Nonlinear.} Taking $d = 2^M$, if we let $\phi_i(\vk)$ be an indicator vector for $\vk$ over all possible bundles in $\mK_i$, then prices are fully nonlinear over bundles.
    \item \textit{Personalized, nonlinear.} In the previous two examples, embeddings were defined so that $\phi_i = \phi_{i'}$ for all agents $i, i' \in N$. To obtain personalized prices, we let each agent $i$ have its own copy of the embedding space. For personalized linear prices, we take $d = nM$ and let $\phi_i(\vk)$ map bundle $\vk$ to a vector that is zero except in agent $i$'s $M$-dimensional subspace, where it contains the item quantities, and similarly for personalized nonlinear prices.
\end{itemize}

With a slight abuse of notation, we define the embedding $\phi$ for an allocation $\vK = (\vk_1, \dots, \vk_n)$ as the sum of the embeddings of the individual bundles, i.e., $\phi(\vK) = \sum_{i \in N} \phi_i(\vk_i)$. This leads to concrete descriptions of the production set $\mK$ in specific applications. For example, in multi-unit auctions (i.e., homogeneous goods) with $K$ units, if we let $\phi_i(\vk)$ be the number of units in bundle $\vk$, then the production set is $\mK = \{\vK : \phi(\vK) \leq K\}$.

\subsection{Multi-Unit and Product-Mix Settings}
\label{sec:multi-productmix}

Our main results provide LP characterizations of UCE prices in full generality, for any choice of embedding $\phi$, but the associated primal-dual auction is necessarily abstract. To demonstrate that our framework truly leads to new auction designs, we will consider special cases in which there are multiple units of each item, but the seller can produce any combination of items as long as the total number of units is at most $K$. This is enough to develop novel iterative Vickrey auctions for two important special cases: multi-unit and product-mix auctions. We first describe the latter.
\paragraph{Product-Mix Auctions.} In this setting, agents have \emph{constant} marginal values for each item type, and the agents agree on the ranking of the items. Specifically, agent $i$'s valuation is determined by values $(v_{i1}, \dots, v_{iM})$, where $v_{i1} \geq v_{i2} \geq \dots \geq v_{iM}$, along with a quantity constraint $\capacity_i$. If $v_{im} = 0$, this implies a constraint that the agent cannot be allocated units of item $m$. Agent $i$'s consumption set is therefore
\[
\mK_i = \left\{ (k_1, \dots, k_M) : \sum_{m=1}^M k_m \leq \capacity_i,\, \mbox{$k_m = 0$ if $v_{im} = 0$} \right\}.
\]
The agent's valuation for $\vk \in \mK_i$ is simply $v_i(\vk) = \sum_{m=1}^M v_{im} k_m$. We let $|\vk| = \sum_{m=1}^M k_m$ be the total number of units in bundle $\vk$.
On the seller's side, the feasibility constraint for allocations is $\sum_{i \in N} |\vk_i| \leq K$. This means that on the seller's part, there is no distinction between different item types. Because agents agree that item 1 has the highest value, there is an efficient allocation that only allocates units of this item. Following~\citet{klemperer2010product}, we will later introduce price differences between items as parameters to create a `supply curve' that will result in a nontrivial space of efficient allocations, but we defer this aspect until Section~\ref{sec:primal-dual-auction} where our primal-dual auction is constructed.

\paragraph{Multi-Unit Auctions.} In this setting there is just a single item type. A bundle is just a single quantity $k$ of units. Marginal values are now \emph{decreasing} rather than constant: for each $i \in N$ we have $v_i(0) = 0$ and $v_i(k+1) - v_i(k) \leq v_i(k) - v_i(k-1)$ for $k > 0$. The agent's consumption set is $\mK_i = \{k : v_i(k) - v_i(k-1) > 0 \; \mbox{or $k = 0$}\}$, meaning it should not be allocated units with zero marginal value. \citet{ausubel2004efficient} has developed an ascending-price \emph{clinching auction} for the multi-unit setting which implements the VCG outcome. Our framework will provide a systematic approach for developing an alternative iterative Vickrey auction via the primal-dual algorithm.

\section{Linear Programming Formulations}
\label{sec:lp-formulations}

In this section we first formulate an LP for the efficient allocation problem, whose dual characterizes CE prices, using the embedding formalism of the previous section to flexibly capture different bundle representations and price structures. We then show how to convert this dual to an LP that characterizes UCE prices.

\subsection{Competitive Equilibrium}

The primal LP computes an efficient allocation. For each $i \in N$ and $\vk \in \mK_i$, we introduce a variable $x_i(\vk) \in [0, 1]$ to capture whether the agent obtains bundle $\vk$ in the allocation. The LP shows the associated dual variables in brackets to the left of each constraint.
\begin{alignat}{3}
    \max_{x \geq 0, y \geq 0} & \qquad \ds \sum_{i \in N} \sum_{\vk \in \mK_i} v_i(\vk) x_i(\vk) && \qquad && \label{opt:primal-ce} \\
    \mbox{$[p]$} & \qquad \ds \sum_{i \in N} \sum_{\vk \in \mK_i} \phi_i(\vk)\, x_i(\vk) = \sum_{\vK \in \mK} \phi(\vK) y(\vK) && && \tag{\ref{opt:primal-ce}a} \label{opt:primal-ce-const-a} \\
    \mbox{$[\pi_i]$} & \qquad \ds \sum_{\vk \in \mK_i} x_i(\vk) \leq 1 && \qquad \mbox{$\forall i \in N$} && \tag{\ref{opt:primal-ce}b} \label{opt:primal-ce-const-b} \\
    \mbox{$[\mu]$} & \qquad \ds \sum_{\vK \in \mK} y(\vK) \leq 1 && && \tag{\ref{opt:primal-ce}c} \label{opt:primal-ce-const-c}
\end{alignat}
Note that constraints~\eqref{opt:primal-ce-const-a} correspond to $d$ different constraints, for each embedding dimension. The dual of this LP takes the following form. There is a variable $\pi_i \geq 0$ for the demand-side constraint of each agent $i \in N$, a variable $\mu \geq 0$ for the supply-side constraint, and a variable $p \in \reals^d$ for the constraints relating demand and supply.
\begin{alignat}{3}
\min_{\pi \geq 0, \mu\geq 0, p\: \mathrm{free}} & \qquad \ds \sum_{i \in N} \pi_i + \mu && \qquad && \label{opt:dual-ce} \\
\mbox{$[x_i(\vk)]$} & \qquad \ds \pi_i \geq v_i(\vk) - p \cdot \phi_i(\vk) && \qquad \mbox{$\forall i \in N$, $\forall \vk \in \mK_i$} && \tag{\ref{opt:dual-ce}a} \label{opt:dual-ce-cons-a} \\
\mbox{$[y(\vK)]$} & \qquad \ds \mu \geq p \cdot \phi(\vK) && \qquad \mbox{$\forall \vK \in \mK$} && \tag{\ref{opt:dual-ce}b} \label{opt:dual-ce-cons-b}
\end{alignat}

\paragraph{Integer solutions.}
The variables in the primal LP are meant to be 0-1 indicator variables for the bundles that agents receive (the $x_i(\vk)$ variables) and the overall allocation (the $y(\vK)$ variables). However, because this is an LP, the variables are continuous and it is possible for the optimal solution to be fractional. To obtain a valid auction with indivisible items, we must ensure that the primal LP has an integer solution by choosing suitable embeddings $\phi_i$. This depends on the structure of the agent valuations $v_i$, and so it is necessarily application-specific. Several characterizations are available in the literature. In multi-unit (homogeneous goods) settings where valuations have decreasing marginal values, one can take $\phi_i(\vk)$ to be the number of units in bundle $\vk$ (here, $d = 1$). In heterogeneous goods settings where agents have substitutes valuations, one can take $\phi_i(\vk)$ to be the 0-1 vector indicating which of the items bundle $\vk$ contains (here, $d = M$); integrality of the resulting LP is nontrivial in this case and relies on the theory of discrete convexity~\citep{murota2016discrete,baldwin2019understanding}. If valuations are fully general, then we must resort to fully personalized, nonlinear embeddings to guarantee integrality~\citep{mishra2007ascending,bikhchandani2002package}.

\medskip
If the primal LP has an integer optimal solution, then an optimal solution $p^*$ to the dual LP corresponds to CE prices, by standard duality arguments~\citep{bikhchandani2002package, bikhchandani2002linear}. We explain the proof here as we will use similar arguments for our UCE results.%
Suppose $x^*$ and $y^*$ are an optimal integer primal solution and $\pi^*$, $\mu^*$, $p^*$ an optimal dual solution. If $x_i^*(\vk^*) = 1$, then the primal LP has computed an efficient allocation where agent $i$ obtains bundle $\vk^*$. By complementary slackness, constraint~\eqref{opt:dual-ce-cons-a} binds and so we have $v_i(\vk^*) - p^* \cdot \phi_i(\vk^*) = \pi_i^*  \geq v_i(\vk) - p^* \cdot \phi_i(\vk)$ for all $\vk \in \mK_i$. In other words, $\vk^* \in D_i(p^*)$ for each $i \in N$. By an analogous appeal to complementary slackness for constraints~\eqref{opt:dual-ce-cons-b}, we have that the efficient allocation maximizes the seller's revenue at prices $p^*$. Thus, $p^*$ supports the efficient allocation computed by the primal LP, represented by the variables $x_i^*(\vk)$ that are 1, or equivalently the variable $y^*(\vK)$ that is 1.

In the sequel we will use $\Primal(N)$ to refer to the primal LP and $\Dual(N)$ for the dual LP, both parametrized by the economy. We write $V(N)$ for their optimal value, which is common to both by strong duality.

\subsection{Universal Competitive Equilibrium}
\label{sec:uce-lp}

We now present our dual LP characterizing UCE prices, referred to as $\UCEDual(N)$.
\begin{alignat}{3}
\min_{\pi \geq 0, \mu \geq 0, \alpha \geq 0, p\: \mathrm{free}, \rho\: \mathrm{free}} & \qquad \ds \sum_{j \in N_0} \left[ \sum_{i \in \Nj} \pi_i^{-j} + \mu^{-j} \right] && \qquad && \label{opt:dual-uce} \\
\mbox{$[x_i^{-j}(\vk)]$} & \qquad \ds \pi_i^{-j} \geq v_i(\vk) - \rho_i(\vk) && \qquad \mbox{$\forall j \in N_0$, $\forall i \in \Nj$, $\forall \vk \in \mK_i$} && \tag{\ref{opt:dual-uce}a} \label{opt:dual-uce-cons-a} \\
\mbox{$[\chi_i^{-j}(\vk)]$} & \qquad \ds \rho_i(\vk) \leq p^{-j} \cdot \phi_i(\vk) + \alpha_i^{-j} && \qquad \mbox{$\forall j \in N_0$, $\forall i \in \Nj$, $\forall \vk \in \mK_i$} && \tag{\ref{opt:dual-uce}b} \label{opt:dual-uce-cons-b} \\
\mbox{$[y^{-j}(\vK)]$} & \qquad \ds \mu^{-j} \geq p^{-j} \cdot \phi(\vK) + \sum_{i \in \Nj} \alpha_i^{-j} && \qquad \mbox{$\forall j \in N_0$, $\forall \vK \in \mK$} && \tag{\ref{opt:dual-uce}c} \label{opt:dual-uce-cons-c}
\end{alignat}
\medskip
To obtain this LP, we applied the following transformation recipe:
\begin{enumerate}
    \item We create copies $\pi_i^{-j}$, $\mu^{-j}$, $p^{-j}$ of the original variables $\pi_i$, $\mu$, $p$ for each economy $j \in N_0$.
    \item We also create new variables $\alpha_i^{-j}$ for all $j \in N_0$, $i \in N^{-j}$.
    \item The objective function is the sum of the dual objectives for the main and marginal economies.
    \item We create copies of the constraints~\eqref{opt:dual-ce-cons-a} for each economy $j \in N_0$, leading to constraint~\eqref{opt:dual-uce-cons-a}. However, each price term $p^{-j} \cdot \phi_i(\vk)$ in a constraint associated to agent $i$ is replaced with a new price variable $\rho_i(\vk)$.
    \item We introduce the new constraints~\eqref{opt:dual-uce-cons-b}. To understand these constraints, the goal is to enforce the relationship:
    $\rho_i(\vk) = \min_{j \in N_0^{-i}} p^{-j} \cdot \phi_i(\vk)  + \alpha_i^{-j}.$
    In other words, $\rho_i(\vk)$ should be the \emph{lower envelope} of the prices for $\vk$ in the main and marginal economies, with additive offsets $\alpha_i^{-j}$ in these economies.
    As the optimization will push $\rho_i(\vk)$ to be as large as possible, this will hold at an optimal solution due to constraints~\eqref{opt:dual-uce-cons-b}.
    \item We create copies of the constraints~\eqref{opt:dual-ce-cons-b} for each economy $j \in N_0$, leading to constraint~\eqref{opt:dual-uce-cons-c}, but these constraints also include the sum of offsets $\sum_{i \in \Nj} \alpha_i^{-j}$ as part of the revenue.
    
\end{enumerate}

The corresponding primal LP is as follows, referred to as $\UCEPrimal(N)$. The common optimal value of the UCE primal and dual is denoted $\tilde{V}(N)$.
\vspace{-10pt}\begin{alignat}{3}
    \max_{x \geq 0, \chi \geq 0, y \geq 0} & \qquad \ds \sum_{j \in N_0} \left[ \sum_{i \in \Nj} \sum_{\vk \in \mK_i} v_i(\vk) x_i^{-j}(\vk) \right] && \qquad &&  \label{opt:primal-uce} \\
    \mbox{$[\pi_i^{-j}]$} & \qquad \ds \sum_{\vk \in \mK_i} x_i^{-j}(\vk) \leq 1 && \qquad \mbox{$\forall j \in N_0$, $\forall i \in \Nj$} &&  \tag{\ref{opt:primal-uce}a} \label{opt:primal-uce-cons-a} \\
    \mbox{$[\rho_i(\vk)]$} & \qquad \ds \sum_{j \in N_0^{-i}} x_i^{-j}(\vk) = \sum_{j \in N_0^{-i}} \chi_i^{-j}(\vk) && \qquad \mbox{$\forall i \in N$, $\forall \vk \in \mK_i$} && \tag{\ref{opt:primal-uce}b} \label{opt:primal-uce-cons-b} \\
    \mbox{$[\alpha_i^{-j}]$} & \qquad \ds \sum_{\vk \in \mK_i} \chi_i^{-j}(\vk) \leq \sum_{\vK \in \mK} y^{-j}(\vK) && \qquad \mbox{$\forall j \in N_0$, $\forall i \in \Nj$} && \tag{\ref{opt:primal-uce}c} \label{opt:primal-uce-cons-c} \\
    \mbox{$[p^{-j}]$} & \qquad \ds \sum_{i \in N} \sum_{\vk \in \mK_i} \phi_i(\vk)\, \chi_i^{-j}(\vk) = \sum_{\vK \in \mK} \phi(\vK) y^{-j}(\vK) && \qquad \mbox{$\forall j \in N_0$} && \tag{\ref{opt:primal-uce}d} \label{opt:primal-uce-cons-d} \\
    \mbox{$[\mu^{-j}]$} & \qquad \ds \sum_{\vK \in \mK} y^{-j}(\vK) \leq 1 && \qquad \mbox{$\forall j \in N_0$} && \tag{\ref{opt:primal-uce}e} \label{opt:primal-uce-cons-e}
\end{alignat}

This LP formulation follows a similar pattern as the dual. Comparing to the original CE primal, we now have a copy $x_i^{-j}(\vk)$ of the original allocation variable $x_i(\vk)$ for each economy $j \in N_0$, but also new allocation variables $\chi_i^{-j}(\vk)$. Constraints~\eqref{opt:primal-uce-cons-d} that balance supply and demand are imposed on the $\chi_i^{-j}(\vk)$ variables, while the $x_i^{-j}(\vk)$ variables appear in the objective. Constraints~\eqref{opt:primal-uce-cons-b} state that, for each agent $i \in N$, the aggregate allocations across economies that contain agent $i$ should equalize.

\subsection{UCE Characterization}

\begin{lemma}
\label{lem:primal-uce-solution}
Assume that the primal $\Primal(\Nj)$ has an integer optimal solution for all $j \in N_0$. Then the primal $\UCEPrimal(N)$ has an integer optimal solution, such that the $x^{-j}_i(\vk)$ variables define an efficient allocation for economy $\Nj$ for each $j \in N_0$, and its optimal value is $\tilde{V}(N) = \sum_{j \in N_0} V(\Nj)$.
\end{lemma}
%
\begin{proof}
We argue that we can assume that at an optimal solution, we have $x^{-j}_i(\vk) = \chi^{-j}_i(\vk)$ for all $j \in N_0$, $i \in \Nj$, $\vk \in \mK_i$. To see this, note that reassigning $x^{-j}_i(\vk) \leftarrow \chi^{-j}_i(\vk)$ obviously maintains equality in constraints~\eqref{opt:primal-uce-cons-b}. As the $x^{-j}_i(\vk)$ values on the left-hand side of~\eqref{opt:primal-uce-cons-b} all have the same coefficient $v_i(\vk)$ in the objective for a fixed $i \in N$ and $\vk \in \mK_i$, this does not change the objective value of the solution. Furthermore, constraints~\eqref{opt:primal-uce-cons-c} and~\eqref{opt:primal-uce-cons-e} imply that the new $x^{-j}_i(\vk)$ values satisfy~\eqref{opt:primal-uce-cons-a}.

Thus, we can introduce the constraints $x^{-j}_i(\vk) = \chi^{-j}_i(\vk)$ for all $j \in N_0$, $i \in \Nj$, $\vk \in \mK_i$ without affecting the optimal value of $\UCEPrimal(N)$. We these extra constraints, the LP simplifies to the following.
\begin{alignat}{3}
    \max_{x \geq 0, \chi \geq 0, y \geq 0} & \qquad \ds \sum_{j \in N_0} \left[ \sum_{i \in \Nj} \sum_{\vk \in \mK_i} v_i(\vk) x_i^{-j}(\vk) \right] && \qquad &&  \nonumber \\
    \mbox{$[\pi_i^{-j}]$} & \qquad \ds \sum_{\vk \in \mK_i} x_i^{-j}(\vk) \leq 1 && \qquad \mbox{$\forall j \in N_0$, $\forall i \in \Nj$} &&  \nonumber \\
    \mbox{$[p^{-j}]$} & \qquad \ds \sum_{i \in N} \sum_{\vk \in \mK_i} \phi_i(\vk)\, x_i^{-j}(\vk) \leq \sum_{\vK \in \mK} \phi(\vK) y^{-j}(\vK) && \qquad \mbox{$\forall j \in N_0$} && \nonumber \\
    \mbox{$[\mu^{-j}]$} & \qquad \ds \sum_{\vK \in \mK} y^{-j}(\vK) \leq 1 && \qquad \mbox{$\forall j \in N_0$} && \nonumber
\end{alignat}
By inspection we see that this LP decomposes into the constraints and variables from $\Primal(\Nj)$ for each $j \in N_0$, and the objective is simply their summed objectives. Thus, as each $\Primal(\Nj)$ has an integer optimal solution, $\UCEPrimal(N)$ also has an integer optimal solution by our previous arguments, such that the variables $x_i^{-j}(\vk)$ define an efficient allocation for economy $\Nj$, for each $j \in N_0$. The optimal value of the summed objective is the sum of the optimal values for $\Primal(\Nj)$ for all $j \in N_0$, or more concisely $\tilde{V}(N) = \sum_{j \in N_0} V(\Nj)$.
\end{proof}

\begin{theorem}
Assume that the primal $\UCEPrimal(N)$ has an integer optimal solution, such that the $x^{-j}_i(\vk)$ variables define an efficient allocation for economy $\Nj$, for each $j \in N_0$. Then at an optimal solution to $\UCEDual(N)$, the $\rho_i(\vk)$ variables represent UCE prices for economy $N$.
\end{theorem}
\begin{proof}
Note that an integer solution to the primal $\UCEPrimal(N)$ must be binary (0 or 1). Fix an economy $j \in N_0$ and let $(\vk^*_1, \dots, \vk^*_n)$ be the efficient allocation defined by the optimal $x_i^{-j}(\vk)$ variables. For each agent $i \in \Nj$ we have
$$
v_i(\vk^*_i) - \rho_i(\vk^*_i) = \pi_i^{-j} \geq v_i(\vk) - \rho_i(\vk)
$$
for all $\vk \in \mK_i$. The first equality holds by complementary slackness, because $x^{-j}_i(\vk^*_i) > 0$, and the remaining inequalities hold by dual feasibility. Thus $\vk^*_i \in D_i(\rho_i)$. Next, let $\vK^*$ be the allocation for which $y^*(\vK^*) = 1$. For all $\vK \in \mK$, we have:
\begin{eqnarray*}
\sum_{i \in \Nj} \rho_i(\vk^*_i) & = & \sum_{i \in \Nj} p^{-j} \cdot \phi_i(\vk^*_i) + \sum_{i \in \Nj}  \alpha_i^{-j} \\
& = & p^{-j} \cdot \phi(\vK^*) + \sum_{i \in \Nj}  \alpha_i^{-j} \\
& = & \mu \\
& \geq & p^{-j} \cdot \phi(\vK) + \sum_{i \in \Nj}  \alpha_i^{-j} \\
& \geq & \sum_{i \in \Nj} \rho_i(\vk_i)
\end{eqnarray*}
The first equality follows from complementary slackness for~\eqref{opt:dual-uce-cons-b}, because we can assume $\chi_i^{-j}(\vk_i^*) = x_i^{-j}(\vk_i^*) > 0$ as in the proof of Lemma~\ref{lem:primal-uce-solution}.
The second equality follow from primal feasibility~\eqref{opt:primal-uce-cons-d} and linearity in $p^{-j}$.
The third equality follows from complementary slackness for ~\eqref{opt:dual-uce-cons-c}, and the next inequality by feasibility for this same constraint.
The final inequality follow from feasibility constraint~\eqref{opt:dual-uce-cons-b}.
This confirms that $(\vk^*_1, \dots, \vk^*_n) \in S(\boldsymbol{\rho})$.
As economy $j$ was arbitrary, we have shown that prices $\boldsymbol{\rho}$ are competitive equilibrium prices for all economies $j \in N_0$, which completes the proof.
\end{proof}

\section{Primal-Dual Auction}
\label{sec:primal-dual-auction}

With our LP characterization of UCE prices in hand, we will now apply the methodology developed by~\citet{de2007ascending} to construct a novel iterative Vickrey auction for the multi-unit and product-mix settings introduced in Section~\ref{sec:multi-productmix}. Upon convergence, the auction obtains an efficient allocation together with UCE prices, from which we can compute and charge VCG payments using formula~\eqref{eq:uce-vcg-formula}. We carefully work out the primal-dual algorithm applied to our UCE dual LP in Section~\ref{sec:restricted-primal-dual}, and formulate an auction based on the price updates it prescribes in Section~\ref{sec:final-uce-auction}.

The literature on optimization offers several primal-dual algorithms. The relevant one here is the primal-dual algorithm for linear programs as described in Chapter 5 of~\cite{papadimitriou1998combinatorial}, which is outlined in Table~\ref{tab:primal-dual-auction-interp}.
\begin{table}[t]
\centering
\begin{tabular}{@{}lp{0.38\textwidth}p{0.38\textwidth}@{}}
\toprule
& \textbf{Primal-Dual Algorithm} & \textbf{Auction Interpretation} \\
\midrule
(P) & Formulate primal. & Formulate allocation problem. \\
\addlinespace
(D) & Formulate dual and pick initial feasible solution. & Formulate pricing problem and pick initial prices. \\
\addlinespace
(RP) & Solve restricted primal to minimize the slack needed to satisfy complementary slackness. & Solve restricted primal that minimizes the slack needed to match supply with demand. \\
\addlinespace[0.3em]
& \begin{minipage}[t]{0.36\textwidth}\leftskip=1.5em\noindent\llap{$\triangleright$\hspace{0.5em}}If slack is zero, output optimal primal and dual solutions.\end{minipage} & \begin{minipage}[t]{0.36\textwidth}\leftskip=1.5em\noindent\llap{$\triangleright$\hspace{0.5em}}If slack is zero, output allocation and supporting prices.\end{minipage} \\
\addlinespace
(DRP) & Solve dual to the restricted primal to find dual update. & Solve dual to the restricted primal to find a price update. \\
\addlinespace[0.3em]
& \begin{minipage}[t]{0.36\textwidth}\leftskip=1.5em\noindent\llap{$\triangleright$\hspace{0.5em}}Apply the update to the dual solution, with suitable step size.\end{minipage} & \begin{minipage}[t]{0.36\textwidth}\leftskip=1.5em\noindent\llap{$\triangleright$\hspace{0.5em}}Apply the update to the current prices, with suitable increment.\end{minipage} \\
\bottomrule
\end{tabular}
\vspace{0.5em}
\caption{Correspondence between primal-dual algorithm and auction steps.}
\vspace{-2em}
\label{tab:primal-dual-auction-interp}
\end{table}
The primal is an allocation problem, but the algorithm operates on the dual pricing problem. We begin with a feasible solution to the dual. Next, we formulate a \emph{restricted primal} which aims to find a solution that satisfies the complementary slackness conditions with respect to the current dual solution. If this is possible, we have a certificate of optimality. Otherwise, we solve the \emph{dual of the restricted primal}. One can show that a solution to the latter leads to a direction of strict improvement for the current dual solution, using a suitable step size to maintain feasibility.

As the table outlines, the mechanics of the primal-dual algorithm have direct interpretations as auction steps. We start with a feasible dual solution, which is usually zero prices. Next, the information needed to formulate the restricted primal is exactly the demand sets from the agents (and the seller's supply set), which is obtained via bidding. Complementary slackness is equivalent to the CE conditions. If these conditions do not hold, then we can formulate and solve the dual of the restricted primal to find a suitable price increment, and the auction continues.

\subsection{Restricted Primal and Dual}
\label{sec:restricted-primal-dual}

To formulate the primal and dual in multi-unit and product-mix settings, we use the embedding $\phi_i(\vk) = |\vk| = \sum_{m=1}^M k_{m}$ which maps a bundle to the number of units it contains. Under this embedding, the LPs for CE and UCE simplify; due to space constraints, we provide them in Appendix~\ref{app:uce-formulations}. The crux of the primal-dual algorithm is to formulate and solve the restricted primal and its dual.

Let 
$\bindingecon = \argmin_{j \in N_0^{-i}} |\vk| p^{-j} + \alpha_i^{-j}$.
Recalling~\eqref{opt:dual-uce-cons-b}, this is the set of economies that determine the lower envelope price $\rho_i(\vk)$, given the $\vp$ and $\boldsymbol{\alpha}$ part of the solution. (We have suppressed the dependence on $\vp$ and $\boldsymbol{\alpha}$ for conciseness.)
The primal complementary slackness conditions state that $x_i^{-j}(\vk) > 0$ only if $\vk \in D_i(\rho_i)$, and $\chi_i^{-j}(\vk) > 0$ only if $j \in \bindingecon$. We therefore set all other variables in the primal to 0. To check whether the primal is still feasible under these restrictions, we introduce slack variables to minimize. This leads to the following restricted primal LP, where the $s$, $t$ variables are slack variables.

\begin{alignat}{3}
    \max_{x \geq 0, \chi \geq 0, s \geq 0, t \geq 0} & \qquad \ds - \sum_{j \in N_0} \left[ s^{-j} + \sum_{i \in \Nj} (s_i^{-j} + t_i^{-j}) \right] - \sum_{i \in N} \sum_{\vk \in \mK_i} s_i(\vk) && \qquad && \nonumber \\
    \mbox{$[\lambda_i^{-j}]$} & \qquad \ds \sum_{\vk \in D_i(\rho_i)} x_i^{-j}(\vk) + s_i^{-j} = 1 && \qquad \mbox{$\forall j \in N_0$, $\forall i \in \Nj$} && \nonumber \\
    \mbox{$[\nu_i^{-j}]$} & \qquad \ds \sum_{\vk: j \in \bindingecon} \chi_i^{-j}(\vk) + t_i^{-j} = 1 && \qquad \mbox{$\forall j \in N_0$, $\forall i \in \Nj$} && \nonumber \\
    \mbox{$[q^{-j}]$} & \qquad \ds \sum_{i \in \Nj} \sum_{\vk : j \in \bindingecon} |\vk| \, \chi_i(\vk) + s^{-j} = K && \qquad \mbox{$\forall j \in N_0$} && \nonumber \\
    \mbox{$[r_i(\vk)]$} & \qquad \ds \sum_{j \in N_0^{-i} \;:\; \vk \in D_i(\rho_i)} x_i^{-j}(\vk) + s_i(\vk) = \sum_{j \in N_0^{-i} \;:\; j \in \bindingecon } \chi_i^{-j}(\vk) && \qquad \mbox{$\forall i \in N, \forall \vk \in \mK_i$} && \nonumber
\end{alignat}
The LP has slack variable $s^{-j}$ for all $j \in N_0$, $s_i^{-j}$ and $t_i^{-j}$ for all $j \in N_0, i \in N^{-j}$, and $s_i(\vk)$ for all $i \in N, \vk \in \mK_i$.

If the optimal solution to the restricted primal is 0, then there is no slack and the complementary slackness conditions can be satisfied, which certifies that the original feasible dual solution
is optimal. If there is slack, then the dual to this LP will provide information on how to update the feasible solution towards optimality. The \emph{restricted dual} is as follows.
\begin{alignat}{3}
\min_{r, \lambda, \nu, q} & \qquad \ds \sum_{j \in N_0} \left[ \sum_{i \in \Nj} \lambda^{-j}_i + K \, q^{-j} + \sum_{i \in \Nj} \nu_i^{-j} \right] && \qquad && \label{opt:restricted-dual} \\
    \mbox{$[x_i^{-j}(\vk)]$} & \qquad \ds \lambda_i^{-j} \geq -r_i(\vk) && \qquad \mbox{$\forall j \in N_0$, $\forall i \in \Nj$, $\forall \vk \in D_i(\rho_i)$} && \tag{\ref{opt:restricted-dual}a} \label{opt:restricted-dual-cons-a} \\
    \mbox{$[\chi_i^{-j}(\vk)]$} & \qquad \ds \nu_i^{-j} \geq r_i(\vk) - |\vk| \, q^{-j} && \qquad \mbox{$\forall i \in N$, $\forall \vk \in \mK_i$, $\forall j \in \bindingecon$} && \tag{\ref{opt:restricted-dual}b} \label{opt:restricted-dual-cons-b} \\
    \mbox{$[s_i^{-j}]$} & \qquad \ds \lambda_i^{-j} \geq -1 && \qquad \mbox{$\forall j \in N_0$, $\forall i \in \Nj$} && \tag{\ref{opt:restricted-dual}c} \label{opt:restricted-dual-cons-c} \\
    \mbox{$[t_i^{-j}]$} & \qquad \ds \nu_i^{-j} \geq -1 && \qquad \mbox{$\forall j \in N_0$, $\forall i \in \Nj$} && \tag{\ref{opt:restricted-dual}d} \label{opt:restricted-dual-cons-d} \\
    \mbox{$[s^{-j}]$} & \qquad \ds q^{-j} \geq -1 && \qquad \mbox{$\forall j \in N_0$} && \tag{\ref{opt:restricted-dual}e} \label{opt:restricted-dual-cons-e} \\
    \mbox{$[s_i(\vk)]$} & \qquad r_i(\vk) \geq -1 && \qquad \mbox{$\forall i \in N$, $\forall \vk \in \mK_i$} && \tag{\ref{opt:restricted-dual}f} \label{opt:restricted-dual-cons-f}
\end{alignat}
When applying the primal-dual algorithm generically, this restricted dual would be solved to optimality using an LP solver. However, any feasible solution to this LP with negative objective value will in fact yield price updates that progress towards optimality of the main UCE dual, as noted by~\citet{de2007ascending}. Although the restricted dual may look complex, the next result establishes that such a solution is easy to construct.

Given current prices $\boldsymbol{\rho}$, let $\mindemand_i = \min_{\vk \in D_i(\rho_i)} |\vk|$ and $\maxdemand_i = \max_{\vk \in D_i(\rho_i)} |\vk|$ for each $i \in N$. We say that there is \emph{over-demand} in economy $j \in N_0$ if $\sum_{i \in N^{-j}} \mindemand_i > K$, and that there is \emph{under-demand} if $\sum_{i \in N^{-j}} \maxdemand_i < K$.
\begin{theorem}
\label{thm:restricted-dual-solution}
Assume that there is over-demand in economy $j \in N_0$. Then there is a feasible solution to the restricted dual with negative objective value, which has $q^{-j} = 1/K$, $q^{-\ell} = 0$ for $\ell \neq j$, $\nu_i^{-j} = 0$ for $i \in N^{-j}$, and $\nu_i^{-\ell} = \mindemand_i/K$ for $\ell \neq j$ and $i \in N^{-\ell}$

Assume that there is under-demand in economy $j \in N_0$. Then there is a feasible solution to the restricted dual with negative objective value, which has $q^{-j} = -1/K$, $q^{-\ell} = 0$ for $\ell \neq j$, $\nu_i^{-j} = 0$ for $i \in N^{-j}$, and $\nu_i^{-\ell} = -\maxdemand_i/K$ for $\ell \neq j$ and $i \in N^{-\ell}$.
\end{theorem}
\begin{proof}
We prove the result for the case of over-demand. The argument for under-demand is analogous.
Assume that economy $j \in N_0$ has over-demand. We set $r_i(\vk) = \min\{|\vk|, \mindemand_i\}/K$.
We set each $\lambda_i^{-j} = -\mindemand_i/K$ (so the latter does not in fact depend on $j$). We set $q^{-j} = 1/K$ and $q^{-\ell} = 0$ for $\ell \neq j$. We set $\nu_i^{-j} = 0$ for $i \in N^{-j}$, and $\nu_i^{-\ell} = \mindemand_i/K$ for $\ell \neq j$ and $i \in N^{-\ell}$. By construction, constraints~\eqref{opt:restricted-dual-cons-c}--\eqref{opt:restricted-dual-cons-f} are clearly satisfied.
For $\vk \in D_i(\rho_i)$, we have $\lambda_i^{-j} = -\mindemand_i/K = -r_i(\vk)$, the latter because $|\vk| \geq \mindemand_i$, so constraints~\eqref{opt:restricted-dual-cons-a} are satisfied.

For constraints~\eqref{opt:restricted-dual-cons-b} and $\ell \neq j$, we have $\nu_i^{-\ell} = \mindemand_i/K \geq r_i(\vk) - |\vk| q^{-\ell}$ as $q^{-\ell} = 0$ and $r_i(\vk) \leq \mindemand_i/K$. This leaves the case of constraints~\eqref{opt:restricted-dual-cons-b} for economy $j$. 
 If $\vk \in D_i(\rho_i)$, then we must have $|\vk| \geq \mindemand_i$, and therefore $r_i(\vk) - |\vk| \, q^{-j} = \mindemand_i/K - |\vk|/K \leq 0 = \nu_i^{-j}$. If $\vk \not\in D_i(\rho_i)$, then $r_i(\vk) - |\vk| \, q^{-j} = |\vk|/K - |\vk|/K = 0 = \nu_i^{-j}$. This confirms feasibility of the constructed solution.

To see that the solution has negative cost, for economy $\ell \neq j$ we have
$$
\sum_{i \in N^{-\ell}} \lambda^{-\ell}_i + K \, q^{-\ell} + \sum_{i \in N^{-\ell}} \nu_i^{-\ell} = \sum_{i \in N^{-\ell}} -\mindemand_i/K + \sum_{i \in N^{-\ell}} \mindemand_i/K = 0,
$$
while for economy $j$ we have
$$
\sum_{i \in \Nj} \lambda^{-j}_i + K \, q^{-j} + \sum_{i \in \Nj} \nu_i^{-j} = \sum_{i \in \Nj} -\mindemand_i/K + 1 < 0
$$
where the inequality follows from over-demand.   
\end{proof}

Given a feasible solution to the restricted dual with negative value, we use this solution to update our current feasible solution to the UCE dual. Using a step size $\Delta > 0$, the updates are:
\begin{alignat}{6}
    p^{-j} & \leftarrow & \; p^{-j} + \Delta \, q^{-j} & \qquad\qquad &
    \alpha_i^{-j} & \leftarrow & \alpha_i^{-j} + \Delta \, \nu_i^{-j} & \nonumber \\
    \pi_i^{-j} & \leftarrow & \; \pi_i^{-j} + \Delta \, \lambda_i^{-j} & \qquad\qquad &
    \rho_i(\vk) & \leftarrow & \; \rho_i(\vk) + \Delta \, r_i(\vk) & \nonumber
\end{alignat}
The next result leads to the step size that we will use in our final iterative auction formulation in Section~\ref{sec:final-uce-auction}. Its proof is deferred to Appendix~\ref{app:omitted-proofs}
\begin{lemma}
\label{lem:restricted-dual-stepsize}
Assume that all values are integer and that marginal values are decreasing. Updating the feasible solution to the UCE dual using the feasible solution from Theorem~\ref{thm:restricted-dual-solution} to its restricted dual, with a step size of $\Delta = K$, leads to a new feasible solution to the UCE dual with strictly lower objective value.
\end{lemma}
\noindent
Note that in the UCE dual~\eqref{opt:dual-uce}, the optimal values of the $\pi_i^{-j}$ and $\rho_i(\vk)$ variables are uniquely determined once the $p^{-j}$ and $\alpha_i^{-j}$ are fixed. Therefore, it suffices to update just these variables during the auction.

\subsection{Iterative UCE Auction}
\label{sec:final-uce-auction}

\begin{algorithm}[t]
    \SetAlgoNoLine
    \KwIn{Initial price $p_\circ$, price differences $\boldsymbol{\delta} = (\delta_1, \ldots, \delta_M)$ with $\delta_M = 0$.}
    \KwOut{Allocation $\vk_i$ and final prices $\rho_i$ for each agent $i \in N$.}
    Initialize $p^{-j} = p_\circ$ for all $j \in N_0$\;
    Initialize $\alpha_i^{-j} = 0$ for all $j \in N_0$, $i \in N^{-j}$\;
    \While{true}{
        Communicate prices $\rho_i(\vk) = \min_{j \in N_0^{-i}} \sum_{m=1}^M k_m (p^{-j} + \delta_m) + \alpha_i^{-j}$\;
        Obtain demand sets $D_i(\rho_i)$ from each agent $i \in N$\;
        \If{$\sum_{i \in N^{-j}} \mindemand_i \leq K \leq \sum_{i \in N^{-j}} \maxdemand_i$ for all $j \in N_0$}{
            Compute allocation such that $\vk_i \in D_i(\rho_i)$ and $\sum_{i \in N} |\vk_i| = K$\;
            \Return allocation $(\vk_1, \dots, \vk_n)$ and prices $\boldsymbol{\rho}$\;
        }
        Pick a $j \in N_0$ where the balance condition does not hold\;
        \If{$\sum_{i \in N^{-j}} \mindemand_i > K$}{
            $p^{-j} \leftarrow p^{-j} + 1$\;
            $\alpha_i^{-\ell} \leftarrow \alpha_i^{-\ell} + \mindemand_i$ for all $\ell \in N_0^{-j}, i \in N^{-\ell}$\;
        }
        \If{$\sum_{i \in N^{-j}} \maxdemand_i < K$}{
            $p^{-j} \leftarrow p^{-j} - 1$\;
            $\alpha_i^{-\ell} \leftarrow \alpha_i^{-\ell} - \maxdemand_i$ for all $\ell \in N_0^{-j}, i \in N^{-\ell}$\;
        }
    }
    \caption{Iterative UCE Auction}
    \label{alg:uce-auction}
\end{algorithm}

In this section we reformulate the primal-dual algorithm for UCE prices as an iterative auction. 
In the UCE dual~\eqref{opt:dual-uce}, the $p^{-j}$ have natural interpretations as constant per-unit prices in the main and marginal economies, and the $\alpha_i^{-j}$ are additive offsets used to form the lower envelope prices.
According to the restricted dual solution from Theorem~\ref{thm:restricted-dual-solution}, and applying the step size from Lemma~\ref{lem:restricted-dual-stepsize}, the update for an economy $j \in N_0$ where there is over-demand is straightforward: we update its price $p^{-j}$ by 1, and we update the offsets $\alpha_i^{\ell}$ by $\mindemand_i$ for $i \in N$ and $\ell \in N_0^{-j}$. (The updates for under-demand are analogous.)

Recall that our model has $M$ item types. However, the primal-dual algorithm only prescribes a price per unit $p^{-j}$ in each economy that does not distinguish between the item types. 
This is due to the fact that the sell-side constraints~\eqref{opt:primal-uce-cons-d} in the UCE primal do not distinguish between the item types. As a result, they are also interchangeable under the lower envelope prices.
To introduce a bias towards allocating certain item types on the part of the seller, \citet{klemperer2010product} introduces a \emph{price difference} parameter vector that we denote $\boldsymbol{\delta} = (\delta_1, \ldots, \delta_M)$, normalized so that $\delta_M = 0$. In our framework, this can be incorporated by replacing each agent's valuation coefficient $v_i(\vk)$ in the primal with $v_i(\vk) - \boldsymbol{\delta} \cdot \vk$.
In the primal-dual procedure, this changes the definition of demand sets. If the unit price is $p$, the utility of a bundle $\vk$ after substituting in the bias becomes 
$$v_i(\vk) - \boldsymbol{\delta} \cdot \vk - |\vk|p = v_i(\vk) - \sum_{m=1}^M (p+\delta_m) k_m.$$
As a result, the unit prices are still parametrized by $p$ which becomes the unit price for item type $M$, while the unit price for item type $m$ becomes $(p + \delta_m)$. For lower envelope prices, the price differences are introduced to the unit prices in the main and marginal economies that are used to form the envelope.

The resulting iterative UCE auction is presented in Algorithm~\ref{alg:uce-auction}. The fact that this auction converges in a finite number of rounds follows from the convergence guarantees of the primal-dual algorithm.
There are a few features to note about practical implementation. First, it is possible for the lower envelope prices to become unnormalized, meaning the zero bundle has nonzero price. This is still correct, as only the relative prices of bundles matter. However, we can simply normalize the prices at each round with the update $\rho_i(\vk) \leftarrow \rho_i(\vk) - \rho_i(\mathbf{0}) = \rho_i(\vk) - \min_{j \in N_0^{-i}} \alpha_i^{-j}.$ This amounts to an update to the $\alpha_i^{-j}$ terms.

Next, note that we only need agents to report the smallest and largest bundle sizes from their demand sets in order to update prices, not the entire demand sets. The full demand sets are only needed to form a supported (i.e., efficient) allocation upon convergence. 
With many units, demand sets can in principle be combinatorially large, posing a problem with reporting them in practice. Because demand sets are convex (more precisely, the restriction of a convex set to integer quantities), it is in fact sufficient for agents to report the extreme points of their demand sets. For the multi-unit setting, this is just a low and high quantity of units. For an agent in a product-mix auction, this is some subset of $\{(0,0), (0, \capacity_i), (\capacity_i, 0)\}$. Therefore, communicating demand sets is not problematic.

We have not elaborated on the algorithm to compute the final allocation when demand balances supply, but it is straightforward. We can start by first allocating a bundle of largest size from each demand set. If this allocation has more than $K$ items in total, then the balance condition implies that we can remove a unit from some agent while still remaining in its demand set. This process is repeated until balance is achieved.

To communicate the lower envelope prices to agent $i$, the seller sends $p^{-j}$ and $\alpha_i^{-j}$ for $j \in N_0^{-i}$, a total of $2n$ numbers. The total dimensionality of the prices is $n^2 + n + 1$. If the number of units is small relative to the number of agents, this can be simplified. The lower envelope prices can be communicated as curves with decreasing marginal price for units, for each agent, which has dimensionality $nm$.

\section{Applications}
\label{sec:applications}




\subsection{Multi-Unit Auction}
\label{sec:multiUnitAuctionExample}

\begin{figure}
    \includegraphics[width=0.79\textwidth]{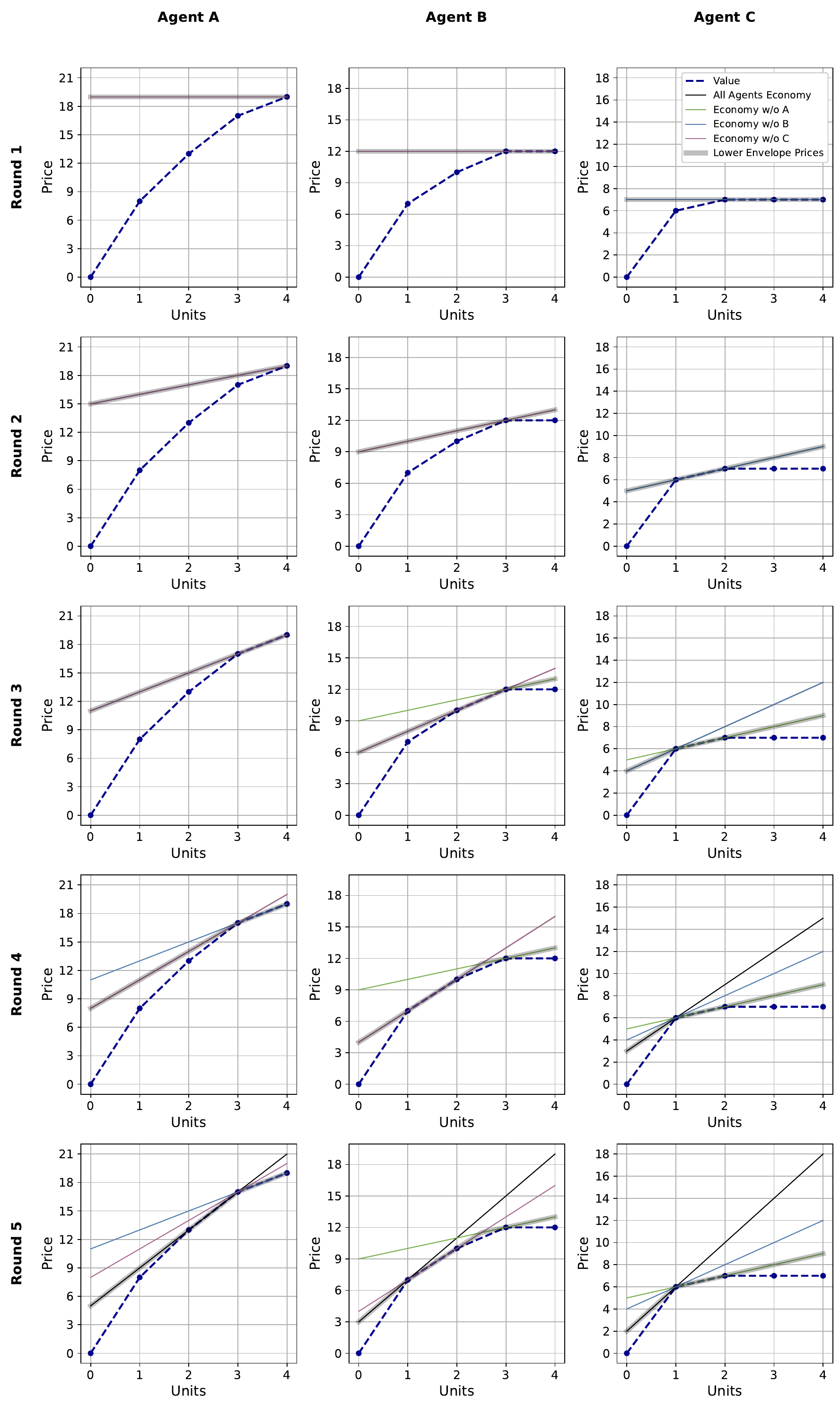}
    \caption{Progression of UCE prices in the multi-unit instance from Table~\ref{tab:marginal-values}. Each panel shows three affine price functions (which sometimes coincide), for the main economy and the marginal economies of the other two agents. Their lower envelope is traced out in gray, and their value with a dashed line.}
    \label{fig:single-item-small}
\end{figure}

We now illustrate the mechanics of our iterative UCE auction in the multi-unit setting. We consider an instance with four units and three agents with decreasing marginal values, shown in Table~\ref{tab:marginal-values}. By inspection, the unique efficient allocation is to give 2 units to A, 1 unit to B, and 1 unit to C. We run the auction from Algorithm~\ref{alg:uce-auction} starting at a unit price of 0. Starting at this low price, there is never under-demand in any of the main or marginal economies during the auction run, so the auction prices are purely ascending.

{
\begin{table}[bht]
    \centering
    \small
    \begin{tabular}{lccc}
    \toprule \toprule
    & Bidder A & Bidder B & Bidder C \\
    \midrule
    Marginal value (1 unit) & 8 & 7 & 6 \\
    Marginal value (2 units) & 5 & 3 & 1 \\
    Marginal value (3 units) & 4 & 2 & 0 \\
    Marginal value (4 units) & 2 & 0 & 0 \\
    \bottomrule
    \end{tabular}
    \medskip
    \caption{Bidder valuations for the worked example of Figure~\ref{fig:single-item-small}.}
    \label{tab:marginal-values}
    \vspace{-15pt}
\end{table}
}

The progression of the auction is shown in Figure~\ref{fig:single-item-small}. For the sake of visualization, prices in each panel are normalized so that a demanded bundle from an agent has utility 0. This means that non-demanded bundles have negative utility, so the prices always dominate the valuation, and they are tangent to the valuation precisely at demanded bundles.

In round 1, prices in all economies are 0. By reading off the smallest bundle size among the tangent points in each panel, we obtain demands (4, 3, 2) for agents (A, B, C) respectively. There is therefore over-demand in the main economy and all marginal economies, so their prices are all increased by one. In round 2, the smallest demand sizes are (4, 3, 1). Now all the economies still have over-demand, except in the economy with A excluded, which clears. In all subsequent panels for agents B and C, the price in this economy (slope of the green line) stays the same. The auction adjusts its intercept so that it stays tangent at the agent's demands from round 2. (It may appear that the intercept does not update in further rounds, but this is due to the renormalization of prices to make them tangent to valuations.)
In round 3, demands are (3, 2, 1). The economy with B excluded has now also cleared, so this price (slope of the blue line) also stays constant going forward. In round 4, the demands are (3, 1, 1), so the economy without C has also cleared. Only the main economy has over-demand, so only its price is incremented. In round 5, the agent demands are (2, 1, 1) so the main economy has finally cleared, and the auction terminates. Recall that this is the efficient allocation.

The way in which the auction updates the additive price offsets when an economy clears is key. The price for the economy freezes, and the offsets are updated in future rounds so that the agent's demanded bundles when the economy clears remain demanded in all future rounds. Taking the lower envelope of the main and marginal economy prices maintains this property, making it possible to use these prices within a single price path. We see from the final round in Figure~\ref{fig:single-item-small} that the valuation difference between the demanded bundles in the main and marginal economies is the same as their price difference.\footnote{Indeed, after normalization the valuations and prices are the same at demanded bundles, but this normalization is only for visualization purposes. The seller does not have enough information, at the end of the auction, to actually normalize prices this way. The normalization does not affect the \emph{difference} between the prices of two bundles.} This is the property that allows one to compute VCG payments from UCE prices via formula~\eqref{eq:uce-vcg-formula}.

\subsection{Stylized Spectrum Auction}

We next illustrate our mechanism on the example introduced in \citet{ausubel2004efficient}, a stylized spectrum auction.  This example includes five agents bidding on multiple units of a spectrum-like asset, with decreasing marginal values.  The efficient allocation is for Agent A to win $3$ items and Agent C to win $2$ items.  The original \citet{ausubel2004efficient} clinching auction implements the Vickrey outcome, where Agent A pays $225$, and Agent C pays $160$ (the other losing agents pay $0$).

Our iterative UCE auction quotes lower envelope prices in each round, with the zero-normalized final round prices for the winners set to $255$ and $170$, respectively.  These prices form a lower-bound on bidder valuations.  We do not require full revelation and thus the seller does not know the available surplus for each winner.  However, we do not implement these prices as final payments.  Instead, we exploit the fact that our price structure contains enough information to compute the Vickrey outcome via formula~\eqref{eq:uce-vcg-formula} and thus the final payments we charge exactly match those of the \citet{ausubel2004efficient} clinching auction.

\begin{figure}
    \centering
    \includegraphics[trim={1.3cm 0 .2cm 0}, clip,width=1\linewidth]{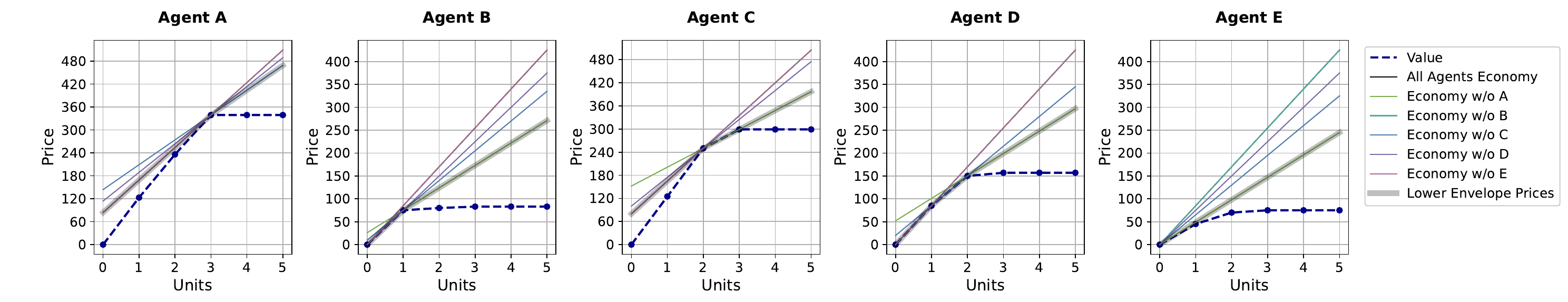}
    \caption{UCE prices in the stylzed spectrum instance from \citet{ausubel2004efficient}. Each panel shows six affine price functions (which sometimes coincide), for the main economy and the marginal economies of the other agents. Their lower envelope is traced out in gray, and their value with a dashed line.}
    \label{fig:ausubel}
\end{figure}

We plot the prices our mechanism produces in this setting in Figure~\ref{fig:ausubel}.  Following the same approach as Figure~\ref{fig:single-item-small}, in Figure~\ref{fig:ausubel} we visualize the prices by re-normalizing them such that the utility for each agents' demanded bundle is $0$.  Concretely, this requires adding $84$ to the prices for Agent A and $80$ to the prices for Agent C.  With this re-normalization it is easy to observe that the prices support the agents' demanded units in all main and marginal economies.

\subsection{Product-Mix Auction}



\begin{table}[b]
   \centering
    \resizebox{\columnwidth}{!}{%
    \addtolength{\tabcolsep}{-2pt}

\begin{tabular}{c|cc|cc|cc|cc}
\toprule
& \multicolumn{2}{c|}{Allocation} & \multicolumn{2}{c|}{Linear Price Auction} & \multicolumn{2}{c|}{Lower Envelope Price Auction} & \multicolumn{2}{c}{$N+1$ Parallel Linear Price Auctions} \\
Price & & & & & & \\
Difference & Strong & Weak & \# Rounds & \# Queries & \# Rounds & \# Queries & \# Rounds & \# Queries \\
\midrule
\multicolumn{9}{c}{Ascending Version} \\
\midrule
13.00 & 0 & 5000 & 21 & 168 & 21 & 168 & 21 & 854 \\
14.00 & 996 & 4004 & 18 & 144 & 18 & 144 & 18 & 662 \\
16.00 & 1191 & 3809 & 4 & 32 & 4 & 32 & 4 & 158 \\
18.00 & 2795 & 2205 & 4 & 32 & 4 & 32 & 4 & 158 \\
19.00 & 3449 & 1551 & 4 & 32 & 4 & 32 & 4 & 158 \\
21.00 & 4315 & 685 & 4 & 32 & 4 & 32 & 4 & 158 \\
22.00 & 5000 & 0 & 4 & 32 & 4 & 32 & 4 & 158 \\
\midrule
\multicolumn{9}{c}{Descending Version} \\
\midrule
13.00 & 0 & 5000 & 1 & 8 & 15 & 120 & 15 & 554 \\
14.00 & 996 & 4004 & 4 & 32 & 18 & 144 & 18 & 746 \\
16.00 & 1191 & 3809 & 18 & 144 & 20 & 160 & 20 & 1250 \\
18.00 & 2795 & 2205 & 18 & 144 & 20 & 160 & 20 & 1250 \\
19.00 & 3449 & 1551 & 18 & 144 & 20 & 160 & 20 & 1250 \\
21.00 & 4315 & 685 & 18 & 144 & 20 & 160 & 20 & 1250 \\
22.00 & 5000 & 0 & 18 & 144 & 20 & 160 & 20 & 1250 \\
\bottomrule
\end{tabular}}
    \caption{Comparison of a linear price auction, our lower envelope price auction, and running $n+1$ linear price auctions in parallel, on a product-mix instance inspired by \citet{grace2024competing}.}
    \label{tab:product_mix}
\end{table}

We next evaluate our iterative UCE auction in the setting of Product-Mix auctions introduced by \citet{klemperer2010product}.  These auctions have been used since 2007 by the Bank of England to allocate loans against both strong and weak collateral as a way to support UK financial institutions, and were especially important in the financial crisis of the Great Recession.  While the Bank of England does not publish the private bidding data from these auctions, Table 1 in a report by \citet{grace2024competing} provides summary statistics about the offer size, number of bidders, bid levels, and allocations in these markets.  We created an instance distributionally consistent with the 3-month term data from this source, details of which are provided in Appendix~\ref{app:ProdMixInstance}.  It has 5000 units for sale (each unit represents 1 million \pounds), 8 bidders each interested in an average of 781 units and an average bids of 2.3bps and 19.6bps for the strong and weak collateral goods respectively.
As described in Section~\ref{sec:final-uce-auction}, the product-mix auction can be parameterized by a price-difference parameter $\delta$ that adjusts the seller's supply curve between strong and weak items. For this specific instance, that parameterization yields $7$ different allocations as $\delta$ is varied.

We compare against two benchmarks. The first benchmark is the conventional iterative linear-price (uniform-price) auction for multi-unit valuations, incorporating a price differential of $\delta$ between weak and strong items. While this auction achieves allocative efficiency (assuming truthful bidding), it does not gather sufficient information to determine VCG payments. The second benchmark executes the linear-price auction simultaneously for both the main and marginal economies, thereby simulating an auction with multiple price trajectories. For this second benchmark, we measure the round count as the maximum over the $n+1$ parallel auctions, while the demand query count represents the total across all sub-auctions.

We implemented ascending versions of the auctions with the strong good initialized at 1.5bps, and descending versions with the strong good starting at 3.5bps. In all runs, the price increment (or decrement) was fixed at 0.1.  We present the results in Table~\ref{tab:product_mix}, with the ascending version listed first, followed by the descending version.  In each case we include one row for each distinct allocation traced by the seller's supply curve. 
For the ascending versions, the three auction types required the same number of rounds. The linear price auction and the lower envelope price auction also used the same number of demand queries, while the parallel linear price auction needed roughly five times as many (i.e., on the order of the number of winning agents), as anticipated. Thus, in the ascending setting, the lower envelope price auction introduces no additional overhead in either rounds or demand queries. This occurs because, by the time the main economy clears, all marginal economies have already cleared.

In the descending versions, when prices arrive at the level that clears the main economy (where the linear price auction terminates), the lower envelope prices have not fallen sufficiently to clear the marginal economies. For the median allocation, the lower envelope price auction incurs an overhead of $11.1\%$ additional rounds and queries compared to the linear price auction. The parallel auction requires the same number of rounds as the lower envelope auction, since it must also clear all main and marginal economies, but its demand queries are larger by a factor of 8.

\section{Conclusion}
\label{sec:conclusion}

This paper has developed a general framework for constructing iterative auctions that implement VCG outcomes using a single price path. The key innovation is a systematic method for transforming linear programs that characterize competitive equilibrium prices into programs that characterize \emph{universal} competitive equilibrium prices, which contain precisely the information needed to compute VCG payments. The transformation preserves important properties like integrality (to handle discrete items), and the resulting lower envelope prices maintain some of the simplicity of the original pricing structure. Applying the primal-dual algorithm to the UCE program leads to iterative auction schemes.

We applied our framework to develop new, single price path VCG auctions for the multi-unit setting and product-mix auctions. An evaluation on an instance of a product-mix auction for Bank of England loans showed that our auction had no overhead, in terms of auction rounds, compared to an ascending linear price auction (which does not compute VCG payments), and minimal overhead in the descending price regime, while dramatically reducing the number of demand queries compared to a multiple price path auction. Notably, the iterative auction implements VCG payments without full information revelation from the bidders.

There are several potential avenues for future work. We intend to fully implement and evaluate our UCE auction for more general valuation classes, including gross substitutes. We believe that the UCE auction’s mechanics can be connected to the concept of clinching in Ausubel’s multi-unit auction, and this could lead to single price path clinching auctions for a broader range of domains. We also intend to investigate whether the UCE transformation preserves monotonicity properties (i.e., ascending or descending prices) of the primal-dual algorithm on the original LP.

\bibliographystyle{plainnat}
\bibliography{iterative-vcg-main}
\vfill
\pagebreak

\appendix
\section{Omitted Proofs}
\label{app:omitted-proofs}

\begin{proof}[Proof of Theorem~\ref{thm:uce-vcg-construction}]
Let $(\vk_1, \dots, \vk_n)$ be an efficient allocation in the main economy and let $(\vk_1^{-\ell}, \dots, \vk_n^{-\ell})$ be efficient allocations in the marginal economies, for $\ell \in N_0$.
By definition, the VCG payment $g_i$ of agent $i$ is:
\begin{align}
  \quad & \ds \sum_{\ell \in N^{-i}} v_{\ell}(\vk_{\ell}^{-i}) -  \sum_{\ell \in N^{-i}} v_{\ell}(\vk_{\ell}) \nonumber \\
= \quad  & \ds \left[ \sum_{\ell \in N^{-i}} v_{\ell}(\vk_{\ell}^{-i}) - \sum_{\ell \in N^{-i}} p_{\ell}(\vk_{\ell}^{-i}) \right] 
   - \left[ \sum_{\ell \in N^{-i}} v_{\ell}(\vk_{\ell}) - \sum_{\ell \in N^{-i}} p_{\ell}(\vk_{\ell}) \right]
   +  \sum_{\ell \in N^{-i}} p_{\ell}(\vk_{\ell}^{-i}) -  \sum_{\ell \in N^{-i}} p_{\ell}(\vk_{\ell}) \nonumber \\
= \quad  & \ds \sum_{\ell \in N^{-i}} \left( [ v_{\ell}(\vk_{\ell}^{-i}) - p_{\ell}(\vk_{\ell}^{-i}) ]
   - [ v_{\ell}(\vk_{\ell}) -  p_{\ell}(\vk_{\ell}) ] \right)
   +  \sum_{\ell \in N^{-i}} p_{\ell}(\vk_{\ell}^{-i}) -  \sum_{\ell \in N^{-i}} p_{\ell}(\vk_{\ell}) \nonumber \\
= \quad  & \ds \sum_{\ell \in N^{-i}} p_{\ell}(\vk_{\ell}^{-i}) -  \sum_{\ell \in N^{-i}} p_{\ell}(\vk_{\ell}) \nonumber
\end{align}
The first two equalities are re-arrangements. The final equality follows because $\vk_i^{-\ell}, \vk_i \in D_i(p_i)$ as they are both supported by the UCE prices, and they therefore have the same utility.
In the final line, the allocations maximize the revenue terms as the UCE prices support them. Therefore, to evaluate these terms, we can use any revenue-maximizing allocations rather than efficient allocations.
\end{proof}

\begin{proof}[Proof of Lemma~\ref{lem:restricted-dual-stepsize}]
The proof of feasibility after applying the update from the restricted dual proceeds by case analysis. For constraints~\eqref{opt:dual-uce-cons-b}, it is routine to check that the updated solution is feasible by considering the cases where 1) $|\vk| \geq \mindemand_i$ versus $|\vk| < \mindemand_i$, and 2) the economy is the selected $j$ which has over-demand, or it is another economy $\ell \neq j$.

For constraints~\eqref{opt:dual-uce-cons-a} and bundles $\vk$ such that $|\vk| \geq \mindemand_i$ , both the left- and right-hand sides are decremented by exactly $\mindemand_i$, so feasibility holds after the update. We therefore consider the case of $|\vk| < \mindemand_i$. Note that this implies $\vk \not\in D_i(\rho_i)$, as $\mindemand_i$ is the minimum size of a demanded bundle. We want to verify the constraint
\begin{equation} \label{eq:constraint-update-feasibility}
\pi_i^{-j} - \mindemand_i \stackrel{?}{\geq} v_i(\vk) - \rho_i(\vk) - |\vk|
\end{equation}
for all $j \in N_0$, $i \in N^{-j}$, $|\vk| < \mindemand_i$. Note that in constraints~\eqref{opt:dual-uce-cons-a}, the $\pi_i^{-j}$ equalize at an optimal solution, so we can drop the superscript. Let $\vk^* \in D_i(\rho_i)$ be such that $|\vk^*| = \mindemand_i$.
We will prove that~\eqref{eq:constraint-update-feasibility} holds by induction on $|\vk^*| - |\vk|$.

For the base case of $|\vk^*| - |\vk|$ = 1, because $\vk^*$ is demanded, $\vk$ is not demanded, and values and prices are integer, we have:
\[
\pi_i = v_i(\vk^*) - \rho_i(\vk^*) \geq v_i(\vk) - \rho_i(\vk) + 1 = v_i(\vk) - \rho_i(\vk) + |\vk^*| - |\vk|, 
\]
which yields~\eqref{eq:constraint-update-feasibility} after re-arranging.

Let $\vk'$ be such that $|\vk'| = |\vk| + 1$. Applying the induction hypothesis, we have for all $\ell \in N_0^{-i}$:
\begin{eqnarray*}
    v_i(\vk^*) - \rho_i(\vk^*) & \geq & v_i(\vk') - \rho_i(\vk') + |\vk^*| - |\vk'|   \\
    & \geq & v_i(\vk') - |\vk'|p^{-\ell} - \alpha_i^{-\ell} + |\vk^*| - |\vk'| \\
    & \geq & v_i(\vk) + p^{-\ell} + 1 - |\vk'|p^{-\ell} - \alpha_i^{-\ell} + |\vk^*| - |\vk'| \\
    & = & v_i(\vk) - |\vk|p^{-\ell}  - \alpha_i^{-\ell} + |\vk^*| - |\vk|
\end{eqnarray*}
The first inequality follows from the induction hypothesis, the second from the fact that $\rho_i$ are lower envelope prices, and the third from decreasing marginal values. Re-arranging $|\vk^*|$ to the left-hand side and taking the maximum of the right-hand side over all $\ell \in N_0^{-i}$ yields~\eqref{eq:constraint-update-feasibility}.
\end{proof}

\section{Linear Programs for Multi-Unit and Product-Mix}
\label{app:uce-formulations}

We first present the primal allocation LP with a supply of $K$ units in total. Using $\phi_i(\vk) = |\vk|$, the formulation simplifies and we can eliminate the $y(\vk)$ variables
\begin{alignat}{3}
    \max_{x \geq 0} & \qquad \ds \sum_{i \in N} \sum_{\vk \in \mK_i} v_i(\vk) x_i(\vk) && \qquad && \nonumber \\
    \mbox{$[p]$} & \qquad \ds \sum_{i \in N} \sum_{\vk \in \mK_i} |\vk|\, x_i(\vk) = K && && \nonumber \\
    \mbox{$[\pi_i]$} & \qquad \ds \sum_{\vk \in \mK_i} x_i(\vk) \leq 1 && \qquad \mbox{$\forall i \in N$} && \nonumber
\end{alignat}
The dual also simplifies, as the optimal revenue is now always $p K$.
\begin{alignat}{3}
\min_{\pi \geq 0, p\: \mathrm{free}} & \qquad \ds \sum_{i \in N} \pi_i + p K && \qquad && \nonumber \\
\mbox{$[x_i(\vk)]$} & \qquad \ds \pi_i \geq v_i(\vk) - p |\vk| && \qquad \mbox{$\forall i \in N$, $\forall \vk \in \mK_i$} && \nonumber
\end{alignat}
Following the recipe from Section~\ref{sec:uce-lp}, we obtain the following LP to characterize UCE prices.
\begin{alignat}{3}
\min_{\pi \geq 0, \alpha \geq 0, p\: \mathrm{free}, \rho\: \mathrm{free}} & \qquad \ds \sum_{j \in N_0} \left[ \sum_{i \in \Nj} \pi_i^{-j} + pK + \sum_{i \in \Nj} \alpha_i^{-j}   \right] && \qquad && \nonumber \\
\mbox{$[x_i^{-j}(\vk)]$} & \qquad \ds \pi_i^{-j} \geq v_i(\vk) - \rho_i(\vk) && \qquad \mbox{$\forall j \in N_0$, $\forall i \in \Nj$, $\forall \vk \in \mK_i$} && \nonumber \\
\mbox{$[\chi_i^{-j}(\vk)]$} & \qquad \ds \rho_i(\vk) \leq p^{-j} |\vk| + \alpha_i^{-j} && \qquad \mbox{$\forall j \in N_0$, $\forall i \in \Nj$, $\forall \vk \in \mK_i$} && \nonumber
\end{alignat}
Finally, we obtain the UCE primal by taking the dual of the preceding LP.
\begin{alignat}{3}
    \max_{x \geq 0, \chi \geq 0} & \qquad \ds \sum_{j \in N_0} \left[ \sum_{i \in \Nj} \sum_{\vk \in \mK_i} v_i(\vk) x_i^{-j}(\vk) \right] && \qquad &&  \nonumber \\
    \mbox{$[\pi_i^{-j}]$} & \qquad \ds \sum_{\vk \in \mK_i} x_i^{-j}(\vk) \leq 1 && \qquad \mbox{$\forall j \in N_0$, $\forall i \in \Nj$} &&  \nonumber \\
    \mbox{$[\rho_i(\vk)]$} & \qquad \ds \sum_{j \in N_0^{-i}} x_i^{-j}(\vk) = \sum_{j \in N_0^{-i}} \chi_i^{-j}(\vk) && \qquad \mbox{$\forall i \in N$, $\forall \vk \in \mK_i$} && \nonumber \\
    \mbox{$[\alpha_i^{-j}]$} & \qquad \ds \sum_{\vk \in \mK_i} \chi_i^{-j}(\vk) \leq 1 && \qquad \mbox{$\forall j \in N_0$, $\forall i \in \Nj$} && \nonumber \\
    \mbox{$[p^{-j}]$} & \qquad \ds \sum_{i \in N} \sum_{\vk \in \mK_i} |\vk|\, \chi_i^{-j}(\vk) = K && \qquad \mbox{$\forall j \in N_0$} && \nonumber
\end{alignat}

\section{Product Mix Auction Instance}
\label{app:ProdMixInstance}

We base our instance on the data in Table 1 of the Bank of England report by \citet{grace2024competing}.  Our instance has 5000 Units for sale, each of which represents a loan for 1 million \pounds.  There are Strong and Weak items for sale, backed by collateral of corresponding quality.  The participants preferred quantities and values are as follows:

\begin{table}[H]
\centering
\begin{tabular}{cccc}
\toprule
Agent & Quantity & Weak & Strong \\
\midrule
A & 911 & 17.2 & 1.6 \\
B & 996 & 16.5 & 3.2 \\
C & 824 & 20.3 & 3.1 \\
D & 685 & 24.1 & 2.2 \\
E & 866 & 22.8 & 1.9 \\
F & 654 & 21.4 & 2.6 \\
G & 780 & 19.7 & 2.4 \\
H & 538 & 15.1 & 1.8 \\
\bottomrule
\end{tabular}
\caption{Bidder preference for an instance of a Product Mix auction for central bank loans.}
\label{tab:ProdMixInstance}
\end{table}

\end{document}